\documentclass[nofootinbib,amsmath,amssymb,amsfonts,aps,prd,eqsecnum,superscriptaddress]{revtex4-1}

\pdfoutput=1
\usepackage{hyperref}
\usepackage{amsmath}
\usepackage{amssymb}
\usepackage{mathrsfs}
\usepackage{xcolor,graphicx}
\usepackage[caption=false]{subfig}
\usepackage[toc,page]{appendix}
\usepackage[section]{placeins}
\usepackage{breqn}
\usepackage[T1]{fontenc}
\usepackage{url}
\usepackage{tabu}
\newcommand{\olsi}[1]{\,\overline{\!{#1}}}

\makeatletter
\let\cat@comma@active\@empty
\makeatother

\begin{document}

\title{Extremely high-order convergence in simulations of relativistic stars}

\author{John Ryan Westernacher-Schneider}
\email{wester5@clemson.edu}
\affiliation{Department of Physics \& Astronomy, Clemson University, \\
118 Kinard Laboratory, Clemson, SC 29634-0978, USA}
\affiliation{Department of Astronomy/Steward Observatory, 
  The University of Arizona, \\
  933 N. Cherry Ave, Tucson, AZ 85721, USA}

\begin{abstract}
We provide a road towards obtaining gravitational waveforms from inspiraling material binaries with an accuracy viable for third-generation gravitational wave detectors, without necessarily advancing computational hardware or massively-parallel software infrastructure. We demonstrate a proof-of-principle $1\! +\! 1$-dimensional numerical implementation that exhibits up to 7th-order convergence for highly dynamic barotropic stars in curved spacetime, and numerical errors up to 6 orders of magnitude smaller than a standard method. Aside from high-order interpolation errors (Runge's phenomenon), there are no obvious fundamental obstacles to obtaining convergence of even higher order. The implementation uses a novel surface-tracking method, where the surface is evolved and high-order accurate boundary conditions are imposed there. Computational memory does not need to be allocated to fluid variables in the vacuum region of spacetime. We anticipate the application of this new method to full $3\! +\! 1$-dimensional simulations of the inspiral phase of compact binary systems with at least one material body. The additional challenge of a deformable surface must be addressed in multiple spatial dimensions, but it is also an opportunity to input more precise surface tension physics.
\end{abstract}

\maketitle

\section{Introduction \& Motivation}\label{sec:intro}
Gravitational wave astronomy has recently exploded with activity~\cite{aasi2015advanced, acernese2014advanced, abbott2016observation, abbott2017gw170817, abbott2019gwtc}. The signal-to-noise ratio in current gravitational wave data~\cite{abbott2017effects, haster2016inference} does not yet challenge the accuracy of theoretical models of the signals coming from perturbation theory, phenomenology, and numerical relativity~\cite{buonanno1999effective, husa2016frequency, boyle2019sxs, dietrich2018core, haas2016simulations, kiuchi2017sub}. Therefore, those models are used to inform interpretations of the data~\cite{abbott2016improved, abbott2019properties}. However, with future third-generation detectors~\cite{abbott2017exploring, reitze2019cosmic} or current detectors at design-sensitivity, the signal-to-noise ratio is expected to challenge current models~\cite{samajdar2018waveform, samajdar2019waveform, brown2020data}. It is therefore necessary to make substantial advancements in the accuracy of theoretical models, including numerical relativity simulations. Advancements are being pursued in perturbative calculations~\cite{goldberger2006effective, galley2009radiation, blanchet2014gravitational, bini2020binary, bini2020sixth, nagar2019efficient, blumlein2020fourth, landry2015tidal, landry2017tidal, pani2015tidal, pani2015tidal2, steinhoff2016dynamical} as well as numerical simulations. Some efforts in the latter category involve innovations and optimizations of hardware~\cite{lim2015technological, stevens2019aurora, narasimhamurthy2019sage}, parallel-computing software~\cite{kidder2017spectre, palenzuela2018simflowny, zhang2019amrex,fernando2018massively, neilsen2019dendro}, numerical methods~\cite{radice2013beyond, miller2016operator, bugner2016solving, bernuzzi2016gravitational, felker2018fourth, foucart2019smooth, most2019beyond, fambri2018ader, fambri2020discontinuous, koppel2018towards, ruchlin2018senr, ruchlin2018senrcode, poudel2020increasing}, and novel physics formulations~\cite{westernacher2019hamilton, westernacher2020exploration}.

In this work, we take advantage of an existing arbitrage opportunity to substantially increase the accuracy of simulations of relativistic stars with a negligible increase in computational cost. The price paid is a modest increase in code complexity. The opportunity exists because current, widely applied numerical methods for dealing with stellar surfaces and vacuum regions are an unnecessary and substantial source of numerical error and degradation of convergence rate. The standard method consists of filling the vacuum regions of spacetime with a artificial low density fluid atmosphere, maintaining it at every time step, and then evolving the fluid equations everywhere (including across stellar surfaces). Often a low-order accurate solver is used purposely in the vicinity of the surface (e.g.~\cite{radice2014high, guercilena2017entropy}). Strong attempts to achieve high-order convergence in binary simulations yield convergence orders for the inspiral gravitational waveform of 2.5-3~\cite{bernuzzi2016gravitational, most2019beyond}.\footnote{Note that Ref.~\cite{bernuzzi2016gravitational} describes oscillations of equilibrium stars in simulations as being triggered by atmosphere effects. However, strictly speaking they are initially triggered by the use of an ill-balanced numerical scheme, whereas atmosphere effects sustain the oscillations subsequently. See e.g.~Ref~\cite{xing2013high} for an example of a well-balanced scheme.} One recent advancement uses negative pressure artificial atmospheres which can be in hydrostatic equilibrium~\cite{westernacher2019hamilton}. A recently demonstrated strategy to use an artificial atmosphere-style scheme with zero density, while importantly lowering errors associated with mass conservation and ejecta, does not substantially improve the bulk motion~\cite{poudel2020increasing}.

We instead evolve the surface position and velocity, and impose free surface boundary conditions there. This moving boundary condition can be imposed at arbitrarily high order, in principle. \emph{We demonstrate up to 7th-order accuracy formally for the fluid variables in a 1+1-dimensional toy model involving highly dynamic stars.} This is presented as a proof-of-principle for the method, with extensions to multiple spatial dimensions left for future work. The multi-dimensional moving boundary (the stellar surface) will also be deformable, which presents an additional technical challenge as well as an opportunity to input surface tension physics. The machinery for dealing with \emph{rigid} moving boundaries in multiple dimensions was already presented in Ref.~\cite{tan2010inverse}. We expect that non-oscillatory interpolators will have to be used in multiple dimensions in order to prevent excessive unphysical wrinkling of the surface, and this may be where surface tension physics enters.

This ``surface-tracking'' method can be used with any interior scheme, which enables adoption of the method in all existing production codes. We anticipate that the method will work with any grid structure, although our demonstration is on a Cartesian one. Furthermore, there is absolutely no need to allocate computational memory to fluid variables in the vacuum regions, although taking advantage of this would increase the data structure complexity in the code.

Our surface boundary treatment is based on a history of development that in the past used the inverse of the original Lax-Wendroff procedure~\cite{lax1960systems}, whereby time derivatives are replaced by spatial derivatives using the equations of motion. In the inverse Lax-Wendroff procedure, spatial derivatives normal to the boundary are replaced by time derivatives and tangential spatial derivatives using the equations of motion. The idea of using the Lax-Wendroff procedure in reverse was proposed in Refs.~\cite{goldberg1978scheme,goldberg1981scheme} for the analysis of numerical boundary conditions, and was extended to arbitrarily shaped boundaries in Ref.~\cite{lombard2008free}. Inspired by Ref.~\cite{huang2008numerical}, extensions to nonlinear conservation laws~\cite{tan2010inverse} and compressible non-relativistic inviscid fluids~\cite{tan2011high} were achieved. Significant simplification of the boundary treatment was presented in Ref.~\cite{tan2012efficient}. Linear stability of the inverse Lax-Wendroff treatment of boundaries has been established for up to 13th-order accurate methods~\cite{vilar2015development, li2016stability}. See the review in Ref.~\cite{shu2017inverse}.

Schemes for tracking a fluid-vacuum interface have previously been reported. A numerical treatment of the vacuum Riemann problem was presented in Refs.~\cite{munz1994tracking, munz1994numerical}. Such work was built upon by Ref.~\cite{tsakiris2007vactrack}, allowing the fluid-vacuum interface to contract, thereby enabling the simulation of oscillating stars with a surface-tracking method. This improved mass conservation relative to an artificial atmosphere scheme, but there was no attempt to render the surface treatment as high-order accurate.

In this work, we extend the work of Ref.~\cite{tan2011high} to relativistic inviscid fluids in 1+1-dimensional curved spacetime, and to boundary conditions appropriate for fluid-vacuum interfaces such as stellar surfaces. The treatment of a fluid-vacuum interface requires novel strategies since extrapolations into the vacuum region would in general yield unphysical values of the fluid variables. We find that we can implement the efficiency improvement of Ref.~\cite{tan2012efficient} even more aggressively, whereby we eliminate the inverse Lax-Wendroff procedure entirely. This yields an exceptionally simple scheme. The reason this is possible is likely due to the fact that stellar surfaces are characteristic surfaces (at least very nearly\footnote{We use a liquid equation of state in this work, whereby the sound speed does not vanish at the stellar surface (but instead becomes very small).}), so one can be more relaxed regarding the treatment of ingoing versus outgoing information.\footnote{If working with an equation of state such that the sound speed is non-negligible at the surface, one can expect to have to use the inverse Lax-Wendroff procedure for the ingoing first derivatives~\cite{tan2012efficient}.} Extension to rigid surfaces in multiple spatial dimensions should follow straightforwardly using the general recipe described in Ref.~\cite{tan2011high}, but as mentioned earlier, treating deformable surfaces will require additional development likely to involve non-oscillatory interpolations of the surface position and velocity. For technical reasons described in the body of the text (see Sec.~\ref{sec:vacbcs}), we use a ``liquid'' equation of state whereby the rest mass density takes on a small positive value when the pressure vanishes (thus the density is finite at the stellar surface). We consider only barotropic fluids, since we anticipate the future application of our methods to the inspiral phase of compact binary mergers involving at least one material body. The inspiral phase is expected to be well-approximated by barotropic fluid flow~\cite{rieutord2006introduction, friedman2013rotating}. The extremeness of the stellar evolutions that we study far exceeds what one expects during the inspiral phase of a binary. In a binary inspiral, stars are subjected to smooth, long-wavelength tidal deformations rather than the volatile sloshing motion that we consider in this work.

It is worth mentioning that the well-posedness of the fluid-vacuum interface is a difficult and unsettled mathematical problem. Some find benefits to using a liquid equation of state (e.g.~Ref.~\cite{oliynyk2019dynamical}). For general reading on the problem, see for example Refs.~\cite{christodoulou2000motion, lindblad2005well, jang2009well, hadvzic2019priori, ginsberg2019priori, oliynyk2019dynamical} and references therein. Numerically, we find that the free surface problem is quite stable when using a standard formulation of hydrodynamics.

In Sec.~\ref{sec:toy}, we describe our toy model environment in which our new method is demonstrated, including the characteristic structure of the system in Sec.~\ref{sec:toyA}. We describe the boundary conditions used at the fluid-vacuum interfaces in Sec.~\ref{sec:vacbcs}, as well our choice of liquid equation of state. The surface-tracking algorithm is described in detail in Sec.~\ref{sec:surftrac}. Implementation details for the interior scheme are presented in Sec.~\ref{sec:imp}. Numerical results are presented in Sec.~\ref{sec:results}. We summarize and give some thoughts on future directions in Sec.~\ref{sec:summary}. Although it is not used, we present the inverse Lax-Wendroff procedure for first derivatives in Appendix~\ref{app:ILW}, as well as the constraint on first derivatives that one would have to use in the inverse Lax-Wendroff procedure in Appendix~\ref{app:surfderconstrainttoy}. Various failure modes that we have encountered for the surface-tracking method, as well as some possible policies for dealing with them, are presented in Appendix~\ref{app:fail} (although we have found that the invocation of such error handling policies usually indicates a bug in the implementation, except for very extreme stellar evolutions).

Throughout, we use the negative-time metric signature $(-,+,+,...)$. Spacetime indices are denoted with letters at the beginning of the alphabet $\lbrace a,b,c... \rbrace$. Spatial indices are denoted with letters beginning in the middle of the alphabet $\lbrace i,j,k... \rbrace$. We use units in which $G=c=1$.

%
%%
%%%
%%%%
%%%
%%
%
\section{Toy Model}\label{sec:toy}
We consider a 1+1-dimensional Cartesian ``star'' in a frozen spacetime with metric
\begin{eqnarray}
ds^2 = -\alpha(x)^2 dt^2 + dx^2. \label{eq:toymetric}
\end{eqnarray}
The spacetime is periodic in $x$ with a length $L$, and the lapse function is prescribed to be
\begin{eqnarray}
\alpha(x) = \frac{2}{3} \left( 1 - \frac{1}{2} \cos\left\lbrace\frac{2 \pi}{L} (x-L/2)\right\rbrace \right). \label{eq:toylapse}
\end{eqnarray}
This lapse function goes to 1 at $x=0$ and $x=L$, and reaches a minimum value of $1/3$ at $x=5$, thereby providing a confining gravitational potential centered at the position $x=5$. In the language of a 1+1-dimensional decomposition of spacetime, the shift vanishes $\beta^x = 0$ and the spatial metric is Euclidean $\gamma_{xx} = \delta_{xx}$, thus in particular the spatial metric determinant is $\sqrt{\gamma}=1$. This Cartesian 1+1-dimensional setup is more representative of a 3+1-dimensional Cartesian setup than a 1+1-dimensional spherically-symmetric one, because of the absence of singular coordinate behavior at the origin. Such singular coordinate behavior in spherical symmetry requires numerical care, and is a distraction from the main proof-of-principle we present in this work. These reasons motivate our choice of toy model spacetime Eq.~\eqref{eq:toymetric}.

We consider two formulations of barotropic fluid dynamics. One resembles the Valencia formulation~\cite{marti1991numerical, font1994multidimensional} because it is based on the equation of momentum balance (we simply refer to it as the Valencia formulation), and the other is based on the Hamiltonian Euler equation (we refer to it as the Hamiltonian formulation). Both formulations use the conservation of rest mass density,
\begin{eqnarray}
0 &=& \nabla_a \left( \rho u^a \right),
\end{eqnarray}
where $\rho$ is the rest mass density and $u^a$ is the fluid 2-velocity. Rest mass conservation is an approximation of baryon number conservation in a relativistic setting. In the usual 1+1 decomposition of the spacetime, this equation becomes
\begin{eqnarray}
0 &=& \partial_t \mathcal{D} + \partial_x \left[ \alpha \mathcal{D} v \right], \label{eq:toyrest} 
\end{eqnarray}
where $v$ is the fluid speed as seen by Eulerian observers, $W = \alpha u^t = 1/\sqrt{1-v^2}$ is the corresponding Lorentz factor, and $\mathcal{D}= \rho W$ is the relativistic rest mass density.

The Valencia formulation uses the covariant conservation of momentum,
\begin{eqnarray}
0 &=& \nabla_a T^a_i,
\end{eqnarray}
where $T^a_i = \rho h u^a u_i + \delta^a_i P$, $h$ is the specific enthalpy, $P$ is the pressure, $a$ is a spacetime index, and $i$ is a spatial index. In a 1+1 decomposition, this equation becomes
\begin{eqnarray}
0 &=& \partial_t S + \partial_x \left[ \alpha S v + \alpha P \right] +  \left(\rho h W^2-P \right) \partial_x \alpha, \label{eq:toyval}
\end{eqnarray}
where $S=\rho h W^2 v$. The Hamiltonian formulation instead uses the Hamiltonian Euler equation
\begin{eqnarray}
0 &=& u^a \left[ \partial_a (h u_b) - \partial_b (h u_a) \right],
\end{eqnarray}
which does not have any covariant derivatives because the derivatives occur in an antisymmetric combination. In the usual 1+1 decomposition of the spacetime, this equation becomes
\begin{eqnarray}
0 &=& \partial_t p + \partial_x H, \label{eq:toyham}
\end{eqnarray}
where $p=hWv$ is the canonical momentum, and $H=\alpha h W$ is the Hamiltonian.\footnote{In an arbitrary frame, stationarity of the fluid configuration implies that a slightly different quantity, namely $\tilde{H}\equiv \alpha h/W$, is constant. This is because stationarity in a moving frame corresponds to vanishing Lagrangian time derivatives $D/Dt = \partial_t + \alpha v \partial_x$. When Eq.~\eqref{eq:toyhamboth} is written in terms of $D/Dt$, stationarity implies that $\alpha h/W$ is constant in space.} See~\cite{westernacher2019hamilton} for a more general discussion of this barotropic Hamiltonian formulation of fluid dynamics.

For ease of reference, we collect the equations of motion for the Valencia formulation here:
\begin{eqnarray}
0 &=& \partial_t \mathcal{D} + \partial_x \left[ \alpha \mathcal{D} v \right], \nonumber\\
0 &=& \partial_t S + \partial_x \left[ \alpha S v + \alpha P \right] +  \partial_x \alpha \left(\rho h W^2-P \right). \label{eq:toyvalboth}
\end{eqnarray}
And we collect the equations of motion for the Hamiltonian formulation here:
\begin{eqnarray}
0 &=& \partial_t \mathcal{D} + \partial_x \left[ \alpha \mathcal{D} v \right], \nonumber\\
0 &=& \partial_t p + \partial_x H. \label{eq:toyhamboth}
\end{eqnarray}
The system is closed by an equation of state, e.g.~$P=P(\rho)$ or $h=h(\rho)$.

\begin{figure}
\centering
\includegraphics[width=0.48\textwidth]{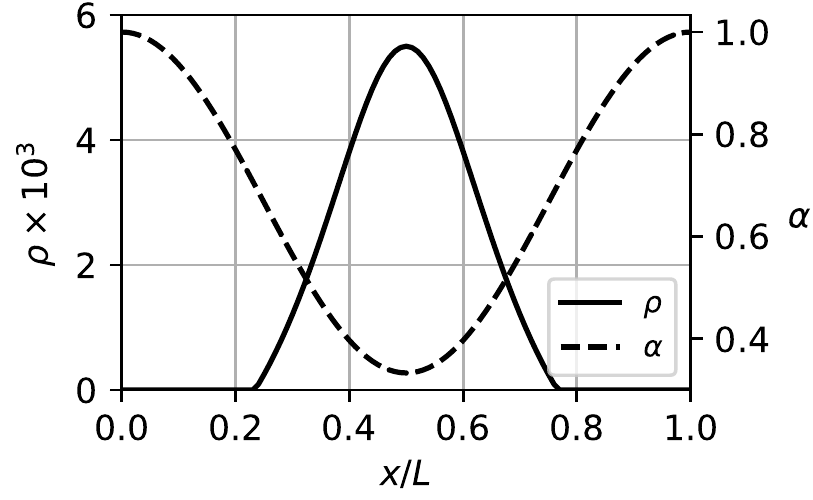}

\caption{Rest mass density for an equilibrium configuration with $H=0.7$, $\kappa=100$, $\Gamma=2$, and lapse function given by Eq.~\eqref{eq:toylapse}.} \label{fig:H07equil}
\end{figure}

A stationary fluid configuration in this spacetime is completely determined by setting $v=0$ and specifying the value of its constant Hamiltonian $H$. The rest mass density profile is then recovered by inverting $H=H(\rho)$. A barotropic equation of state can be written as $h=h(\rho)$. Thus the recovery of $\rho$ from a chosen value of $H$ begins with obtaining $h$ via $h=H/\alpha$, and then inverting $h=h(\rho)$. For the lapse function given in Eq.~\eqref{eq:toylapse}, the fluid has left and right surfaces when $H<1$, which correspond to spatial locations $x$ where $h(x)=1$. When $H\geq 1$, the fluid extends across the entire spatial domain. In Fig.~\ref{fig:H07equil} we display the lapse function and resulting density profile for the choice $H=0.7$, and a polytropic equation of state $P=\kappa \rho^\Gamma$ with $\kappa = 100$, and $\Gamma=2$. This star has a central density $\rho_c = 5.5\times 10^{-3}$, central energy density $\varepsilon_c \approx 8.7\times 10^{-3}$, dynamical time $t_d = 1/\sqrt{\varepsilon_c} \approx 10$, and left and right surfaces at $x/L \approx 0.2341$ and $x/L\approx 0.7659$, respectively. 

We also consider increasingly perturbed versions of this configuration, whereby the star is initialized with uniform motion $u^x/u^t=\alpha v\in \lbrace 0.01,0.02,0.04 \rbrace$. In these cases, the initial enthalpy is given by $h=\tilde{H}W/\alpha$ with $\tilde{H}=0.7$, and the initial surface locations are correspondingly modified (but still satisfy $h=1$). Note that the spacetime remains fixed, so that the resulting evolution is a star that sloshes back and forth in a fixed gravitational well.

Note that we use an equation of state that is not precisely $P=100 \rho^2$, but is very close (and described in Sec.~\ref{sec:vacbcs}).

%
%%
%%%
%%
%
\subsection{Characteristic Structure} \label{sec:toyA}
In this section we present the characteristic structure used in our numerical implementation. For a set of evolved variables $\boldsymbol{U}$, corresponding fluxes $\boldsymbol{F}$, sources $\boldsymbol{S}$, and lapse function $\alpha(x)$, the equations of motion read
\begin{eqnarray}
0 &=& \partial_t \boldsymbol{U} + \partial_x \left[\alpha \boldsymbol{F} \right] + \boldsymbol{S}.
\end{eqnarray}
When seeking the characteristic structure of this system with respect to a chosen set of variables $\boldsymbol{q}$, one can use the chain rule as follows:
\begin{eqnarray}
0 &=& \frac{\partial \boldsymbol{U}}{\partial \boldsymbol{q}} \partial_t \boldsymbol{q} + \alpha \frac{\partial\boldsymbol{F}}{\partial\boldsymbol{q}} \partial_x \boldsymbol{q} + \mathrm{l.o.} \nonumber\\
\rightarrow 0 &=& \partial_t \boldsymbol{q} + \alpha \left( \frac{\partial \boldsymbol{U}}{\partial \boldsymbol{q}} \right)^{-1} \frac{\partial\boldsymbol{F}}{\partial\boldsymbol{q}} \partial_x \boldsymbol{q} + \mathrm{l.o.},
\end{eqnarray}
where we have denoted by $\mathrm{l.o.}$ terms that are lower order in the sense of characteristic analysis,\footnote{For example, terms proportional to $\boldsymbol{S}$ and $\partial_x \alpha$.} and assumed $\partial \boldsymbol{U}/\partial \boldsymbol{q}$ is invertible. This is the quasilinear form of the equations of motion, and the Jacobian is $J \equiv \alpha (\partial\boldsymbol{U}/\partial\boldsymbol{q})^{-1} \partial \boldsymbol{F}/\boldsymbol{q}$. If the fluxes and evolution variables can be written as explicit functions of the choice of variables $\boldsymbol{q}$, then the Jacobian is easy to compute.

We use the characteristic structure associated with two choices of variables: ``safe''\footnote{The name ``safe'' is explained in Sec.~\ref{sec:vacbcs}.} variables $\boldsymbol{q} = (h,Wv)^T$ and ``flux'' variables $\boldsymbol{q} = \boldsymbol{F}$. For the flux variables, the Jacobian is the same as if $\boldsymbol{q}=\boldsymbol{U}$, namely
\begin{eqnarray}
J = \alpha \frac{\partial \boldsymbol{F}}{\partial \boldsymbol{U}},
\end{eqnarray}
which is readily computed in terms of primitive variables $\boldsymbol{p}$ using the chain rule and assuming the applicability of the inverse function theorem for $\partial \boldsymbol{p}/\partial \boldsymbol{U}$: $J = (\partial \boldsymbol{F}/\partial \boldsymbol{p})(\partial \boldsymbol{U}/\partial \boldsymbol{p} )^{-1}$.

For the safe variables, we use the full expression for the Jacobian,
\begin{eqnarray}
J = \alpha \left( \frac{\partial \boldsymbol{U}}{\partial \boldsymbol{q}}\right)^{-1} \frac{\partial \boldsymbol{F}}{\partial \boldsymbol{q}}.
\end{eqnarray}
The matrices $\partial \boldsymbol{U}/\partial \boldsymbol{q}$ and $\partial \boldsymbol{F}/\partial \boldsymbol{q}$ are again readily computed in terms of primitive variables $\boldsymbol{p}$ via $\partial \boldsymbol{U}/\partial \boldsymbol{q} = (\partial \boldsymbol{U}/\partial \boldsymbol{p}) (\partial \boldsymbol{q}/\partial \boldsymbol{p})^{-1}$ and $\partial \boldsymbol{F}/\partial \boldsymbol{q} = (\partial \boldsymbol{F}/\partial \boldsymbol{p}) (\partial \boldsymbol{q}/\partial \boldsymbol{p})^{-1}$, which assumes the applicability of the inverse function theorem for $\partial \boldsymbol{q}/\partial \boldsymbol{p}$.

The Jacobian $J$, its eigenvalues $\lambda_{\pm}$, and its left and right eigenvectors $\boldsymbol{L}_{\pm}$, $\boldsymbol{R}_{\pm}$ are provided in Table~\ref{tab:char} for the safe variables and flux variables in the Valencia and Hamiltonian formulations. The speed of sound is $c_s = \sqrt{(1/h)\partial_{\rho} P}= \sqrt{(\rho/h)\partial_{\rho}h}$. In our numerical implementation we make use of local characteristic variables defined by
\begin{eqnarray}
\boldsymbol{V} =
	\begin{bmatrix}
		L_{+,1} & L_{+,2} \\
	    L_{-,1} & L_{-,2} 
%%%% this for (\rho, Wv) formulation
%	\phantom{+}Wc_s/\rho & 1 \\
%	-Wc_s/\rho & 1 
\end{bmatrix}_{X}
\boldsymbol{q},
\end{eqnarray}
where $\boldsymbol{L}_{\pm}=(L_{\pm,1},L_{\pm,2})$, and the matrix of left eigenvectors is evaluated at the specific point $x=X$, where the characteristic decomposition is being performed.

{\renewcommand{\arraystretch}{2.5}
\begin{table}[]
\begin{tabular}{|l|l|l|}
\hline
Safe variables & Valencia flux variables & Hamiltonian flux variables \\ \hline
\(\displaystyle
J_{\phantom{\pm}}\; = \frac{\alpha}{1-v^2 c_s^2}
  \begin{bmatrix}
     v (1-c_s^2) & h c_s^2/W^3 \\
    1/(h W) & v (1-c_s^2)
  \end{bmatrix}
\)
& 
\(\displaystyle
J_{\phantom{\pm}}\; = \frac{\alpha}{1-v^2 c_s^2}
  \begin{bmatrix}
     v (v^2-c_s^2) & 1/(h W^3) \\
    \frac{h}{W}(v^2-c_s^2) & v(2-v^2-c_s^2)
  \end{bmatrix}
\)
& 
\(\displaystyle
J_{\phantom{\pm}}\; = \frac{\alpha}{1-v^2 c_s^2}
  \begin{bmatrix}
     v (1-c_s^2) & \rho/(hW^2) \\
    h c_s^2/(\rho W^2) & v (1-c_s^2)
  \end{bmatrix}
\)
                        \\
\(\displaystyle \lambda_{\pm}\; = \alpha \frac{v \pm c_s}{1\pm v c_s} \)
&
\(\displaystyle \lambda_{\pm}\; = \alpha \frac{v \pm c_s}{1\pm v c_s} \)
&
\(\displaystyle \lambda_{\pm}\; = \alpha \frac{v \pm c_s}{1\pm v c_s} \)
						\\
\(\displaystyle \boldsymbol{L}_{\pm}\, = (\pm \frac{W}{h c_s},1) \)
&
\(\displaystyle \boldsymbol{L}_{\pm}\, = (hW(-v\pm c_s),1) \)
&
\(\displaystyle \boldsymbol{L}_{\pm}\, = (\pm \frac{h c_s}{\rho},1) \)
						\\
\(\displaystyle \boldsymbol{R}_{\pm} = (\pm \frac{h c_s}{W},1)^T \)
&
\(\displaystyle \boldsymbol{R}_{\pm} = ( \frac{1}{hW(v\pm c_s)},1)^T \)
&
\(\displaystyle \boldsymbol{R}_{\pm} = (\pm \frac{\rho}{h c_s},1)^T \)
%						\\
%\(\displaystyle \mathcal{V}_{\pm}\; = \frac{W}{c_s}(v c_s \pm 1) \)
%&
%\(\displaystyle \mathcal{V}_{\pm}\; = \frac{W}{c_s}(v c_s \pm 1) \)
%&
%\(\displaystyle \mathcal{V}_{\pm}\; = \frac{W}{c_s}(v c_s \pm 1) \)
						\\ \hline
\end{tabular}
\caption{Collection of fluid characteristic information with respect to several sets of variables.} \label{tab:char}
\end{table}
}

%
%%
%%%
%%%%
%%%
%%
%
\section{Vacuum boundary conditions} \label{sec:vacbcs}
For concreteness we focus on the right boundary of our 1-dimensional star. The left boundary description is analogous. The boundary conditions appropriate for the right stellar surface are
\begin{alignat}{3}
P(x_v) &\equiv  \olsi{P} &&= 0, \nonumber\\
v(x_v) &\equiv  \olsi{v} &&= \lim_{x\rightarrow x_v^{-}} v(x). \label{eq:surfbcs}
\end{alignat}
We denote evaluation at the surface with an overbar. The vanishing pressure condition implies $\olsi{h}=1$. We assume a liquid equation of state whereby the rest mass density takes a positive value $\olsi{\rho}$ when the pressure vanishes, i.e. $P(\olsi{\rho})=0$ and $\olsi{\rho}>0$. The use of a liquid equation of state greatly simplifies technical details in the description of the stellar surface, because it allows one to work directly at the surface rather than in a neighborhood of it (essentially by avoiding repeated divisions by zero there). An example of such an equation of state is that of a generalized polytrope, 
\begin{eqnarray}
h = 1 - a + \frac{\kappa \Gamma}{\Gamma-1} \rho^{\Gamma-1}
\end{eqnarray}
with $a=\mathrm{constant}>0$. The corresponding formula for the density is 
\begin{eqnarray}
\rho = \left( \frac{h-1+a}{\kappa (1+n)} \right)^n
\end{eqnarray}
where $\Gamma = 1+1/n$. The pressure vanishes when the enthalpy $h=1$, implying that
\begin{eqnarray}
\olsi{\rho} = \left(\frac{a}{\kappa(1+n)}\right)^n.
\end{eqnarray}
By taking $a\ll 1$, this equation of state deviates negligibly from the usual polytrope except in the vicinity of the surface of a star. In this work, we set $\olsi{\rho}=10^{-13}$, which is roughly 10 orders of magnitude below the central density of the star. We found no issue going even lower, and our results on convergence are not sensitive to the choice. The specific evolutions we obtain are also not sensitive to the choice of $\olsi{\rho}$, provided it is chosen to be sufficiently small (a similar level as with artificial atmosphere prescriptions, i.e. at least 6 or 7 orders of magnitude below the central density of the star).

In past work, it was found that high-order numerical treatment of boundary conditions can be achieved stably as long as the inverse Lax-Wendroff procedure is followed for ingoing first derivatives~\cite{tan2012efficient}, with all ingoing higher derivatives simply being extrapolated to the boundary. For a stellar surface, we find we are able to be even more aggressive; we can extrapolate \emph{all} ingoing derivatives, and only impose the free surface boundary conditions on the undifferentiated fluid variables. This makes the implementation especially efficient. The greater stability we find for the stellar surface problem is probably because the surface is almost a characteristic boundary ($c_s\ll 1$). If the ingoing and outgoing characteristic speeds were very different, extrapolation of ingoing variables to the boundary would be numerically more dangerous. If the sound speed is not small at the surface, one can expect to have to follow the inverse Lax-Wendroff procedure for at least the ingoing first derivatives, as in~\cite{tan2012efficient}. In Appendix~\ref{app:surfderconstrainttoy} we have derived the constraint on first spatial derivatives of the fluid variables at the stellar surface, which would be necessary for the inverse Lax-Wendroff procedure. We also describe how to derive constraints on higher order derivatives.

As in Ref.~\cite{westernacher2019hamilton}, we extend the equation of state to negative pressures $h<1$ via a piecewise attachment to $h(\rho)$. This is why we call the variables $\boldsymbol{q} = (h,Wv)^T$ ``safe,'' i.e. if extrapolations result in a value of the enthalpy less than 1 (but greater than 0), the equation of state supports it. The variable $Wv$ is also safe (compared to $v$) because its physically-admissible range is $(-\infty,\infty)$.

%
%%
%%%
%%%%
%%%
%%
%
\section{Surface tracking} \label{sec:surftrac}
In this section, we describe our high-order accurate tracking algorithm for the fluid-vacuum interface. We follow Ref.~\cite{tan2011high} closely, and adapt their work to the relativistic case and to boundary conditions which are appropriate for the fluid-vacuum interface. We also combine the method with an efficiency improvement presented in~\cite{tan2012efficient}, which we implement even more aggressively. The basis of the original approach is the inverse Lax-Wendroff method, which replaces spatial derivatives at the boundary with time derivatives, and was presented for nonlinear conservation laws originally in~\cite{tan2010inverse}\footnote{Note there is a sign typo in their equation (2.16). It should read $\beta_2 = (61 u_0^2 + 160 u_1^2 + 74 u_0 u_2 + 25 u_2^2 - 196 u_1 u_0 - 124 u_1 u_2)/12$}. But our implementation does not use the inverse Lax-Wendroff method at all, making it simpler to implement. For concreteness and ease of presentation, we use a 5th-order accurate scheme~\cite{tan2012efficient} in what follows, but other orders of accuracy can be obtained with obvious modifications (which we point out along the way).

First, it is instructive to give a few words about the time update procedure. At time $t^{n}$, where the exponent denotes a time level index, we assume the last interior point on the right side of the star is located at position $x_R$. Everything that follows works similarly for the left surface. The surface at time $t^n$ is located at $x_v$ such that $x_{R} \leq x_v < x_{R+1}$, and $x_{R+1}$ is a vacuum point. The surface may or may not cross a grid point during the update (in a sub-step of the update, or after the entire update), either engulfing a new grid point or receding past one. If the surface engulfs the new grid point $x_{R+1}$ in any sub-step of the update, values for the fluid variables are not assigned there until the full time step is complete, and only if the surface position at time $t^{n+1}$ has engulfed the point $x_{R+1}$.\footnote{It is possible that the surface could engulf a grid point or recede past a grid point in a sub-step of the update, but not after the entire update.} If the surface recedes past the grid point $x_R$ during any sub-step or after the entire update, strictly speaking one should anticipate this at the beginning of the time update, and then only evolve the points that are interior to the star during all sub-steps and at time $t^{n+1}$, i.e.~$\lbrace \ldots, x_{R-3}, x_{R-2},x_{R-1} \rbrace$. However, rather than devise such an anticipation criterion, we instead drop the point $x_R$ in the middle of the time update if the surface recedes past it on any sub-step of the time update (e.g.~Runge-Kutta sub-steps), and continue to exclude that point for the remainder of the update. If $x_R$ is still an interior point at time $t^{n+1}$, then we repopulate the fluid variables there at that stage.

The fluid variables at the interior points near the surface, which we collect as $X_{\mathrm{int}} \equiv \{ x_{R-4}, x_{R-3}, x_{R-2}, x_{R-1}, x_R \}$,\footnote{If using the 7th-order scheme, then $X_{\mathrm{int}} = \{ x_{R-6}, x_{R-5}, x_{R-4}, x_{R-3}, x_{R-2}, x_{R-1}, x_R \}$.} are assumed to have been updated (and possibly defined anew for $x_R$) from time $t^{n-1}$ to time $t^n$, or they are provided by initial data if the initialization time is $t^n$. We will be assigning values of the split fluid fluxes $\boldsymbol{f}^{\pm}$ (defined in Sec.~\ref{sec:imp}, and we denote simply as $\boldsymbol{f}$ in this section) at ghost points located at $X_{\mathrm{ghost}}\equiv \lbrace x_{R+1}, x_{R+2} \rbrace$\footnote{If using the 7th-order scheme, then $X_{\mathrm{ghost}} = \lbrace x_{R+1}, x_{R+2}, x_{R+3} \rbrace$} via a Taylor expansion at the fluid-vacuum interface $x_v$:
\begin{eqnarray}
\boldsymbol{f}_i = \sum_{k=0}^4 \frac{(x_i-x_v)^k}{k!} \olsi{\boldsymbol{f}}^{(k)}, \label{eq:taylor}
\end{eqnarray}
where $\olsi{\boldsymbol{f}}^{(k)}$ is the $k$th-order derivative of $\boldsymbol{f}$ approximated at $(5-k)$th order\footnote{$(7-k)$th order for a 7th-order scheme.} (denoted with the superscript$\,^{(k)}$) evaluated at the surface (denoted with the overbar).

We transform to \emph{local} characteristic variables at the point $x_R$ in order to control the outgoing versus ingoing fluid information. This is done separately for the unsplit fluxes $\boldsymbol{F}$ and the safe variables $\boldsymbol{q}$ from Sec.~\ref{sec:toy}. Using the characteristic structure from Table~\ref{tab:char}, the local characteristic variable transformation for any set of variables (represented schematically as $\boldsymbol{Q}$) is
\begin{eqnarray}
^{\boldsymbol{Q}}\boldsymbol{V}_{i} =
	\begin{bmatrix}
		L_{+,1} & L_{+,2} \\
	L_{-,1} & L_{-,2}
%%%% this for (\rho, Wv) formulation
%	\phantom{+}Wc_s/\rho & 1 \\
%	-Wc_s/\rho & 1 
\end{bmatrix}_{x_R}
\boldsymbol{Q}_i. \label{eq:locchar0}
\end{eqnarray}
Notice that the matrix of left eigenvectors is evaluated at the point $x_R$, so by taking the derivative of both sides, we see that derivatives of the variables also transform in the same way:
\begin{eqnarray}
^{\boldsymbol{Q}}\boldsymbol{V}^{(k)}_{i} =
	\begin{bmatrix}
		L_{+,1} & L_{+,2} \\
	L_{-,1} & L_{-,2}
%%%% this for (\rho, Wv) formulation
%	\phantom{+}Wc_s/\rho & 1 \\
%	-Wc_s/\rho & 1 
\end{bmatrix}_{x_R}
\boldsymbol{Q}^{(k)}_i. \label{eq:locchark}
\end{eqnarray}

After obtaining local characteristic variables corresponding to the unsplit fluxes $\boldsymbol{F}$ and the safe variables $\boldsymbol{q}$, we fit a 5th-order Lagrange polynomial\footnote{7th-order Lagrange polynomial for a 7th-order scheme} to the available grid information near the surface, at locations $X_{\mathrm{int}}$. Derivatives approximated at $(5-k)$th order (up to $k=4$),\footnote{$(7-k)$th order up to $k=6$ for a 7th-order scheme.} $^{\boldsymbol{Q}}\boldsymbol{V}^{(k)}_i$, are obtained by analytic derivatives of the Lagrange polynomial fits.
If the solution near the surface is expected to be non-smooth, a weighted essentially non-oscillatory (WENO) polynomial may be more appropriate~\cite{tan2012efficient}. But we find that the bare Lagrange polynomials work quite well.\footnote{On physical grounds, fluid-vacuum interfaces do not technically support shockwaves because the Rankine-Hugoniot jump conditions are not satisfied there~\cite{munz1994tracking, toro2013riemann}. But that fact does not exclude the possibility of gradients near the surface that are so large as to present a practical numerical problem for simple Lagrange polynomial extrapolation.} In addition to denoting evaluation at the surface with an overbar, e.g.~$\olsi{\boldsymbol{Q}}$, we denote extrapolated values with a star superscript, e.g.~$\boldsymbol{Q}^{*}$.

We extrapolate all the local characteristic variables to the surface, thereby obtaining $^{\boldsymbol{q}}\olsi{\boldsymbol{V}}^{(k)}$ and $^{\boldsymbol{F}}\olsi{\boldsymbol{V}}^{(k)}$. The only exceptions are the ingoing variables for $k=0$, $^{\boldsymbol{q}}\olsi{V}^{(0)}_{-}$ and $^{\boldsymbol{F}}\olsi{V}^{(0)}_{-}$, which are instead effectively fixed by the free surface boundary condition $\rho\vert_{x_v} = \olsi{\rho}$. For example, we solve for $\olsi{v}$ using $^{\boldsymbol{q}}\olsi{V}^{*}_{+}$ and using Eq.~\eqref{eq:locchar0} evaluated at the surface:
\begin{eqnarray}
^{\boldsymbol{q}}\olsi{V}^{*}_{+} = \left(\frac{W}{h c_s}\right)_{\!\! R} + \olsi{W} \olsi{v}. \label{eq:surf0}
\end{eqnarray}
In this expression, the only unknown is $\bar{v}$ and one can always solve for it. Thus, we obtain $\olsi{v}$ from Eq.~\eqref{eq:surf0}, and $\olsi{\rho}$ is the (constant) surface density value determined by the liquid equation of state. This yields $\olsi{\boldsymbol{q}}^{(0)}$, from which $\olsi{\boldsymbol{F}}^{(0)}$ can easily be computed.

In this way, we obtain $\olsi{\boldsymbol{q}}^{(k)}$ and $\olsi{\boldsymbol{F}}^{(k)}$, from which we form the split fluxes $\olsi{\boldsymbol{f}}_{\pm}^{(k)}$ from Sec~\ref{sec:imp}. Flux values are then obtained at any points where they are needed using the Taylor expansion Eq.~\eqref{eq:taylor}. If populating fluid variables at interior points, we do so with the equivalent Taylor expansion for the safe variables $\boldsymbol{q}$.

As per the style in Refs.~\cite{tan2010inverse,tan2011high, tan2012efficient}, we now summarize the procedure for populating fluxes at ghost points and possibly populating fluid variables at interior points:
\begin{enumerate}
	\item Compute the local characteristic variables $^{\boldsymbol{q}}\boldsymbol{V}_i$ and $^{\boldsymbol{F}}\boldsymbol{V}_i$ at points $X_{\mathrm{int}}$ using Eq.~\eqref{eq:locchar0}.
	\item Fit Lagrange polynomials to $^{\boldsymbol{q}}\boldsymbol{V}_i$ and $^{\boldsymbol{F}}\boldsymbol{V}_i$ at points $X_{\mathrm{int}}$, then, using those polynomials, compute derivatives $^{\boldsymbol{q}}\boldsymbol{V}^{(k)}$ and $^{\boldsymbol{F}}\boldsymbol{V}^{(k)}$ and extrapolate to the surface to obtain $^{\boldsymbol{q}}\olsi{\boldsymbol{V}}^{*(k)}$ and $^{\boldsymbol{F}}\olsi{\boldsymbol{V}}^{*(k)}$.
	\item Obtain $\olsi{\boldsymbol{q}}^{(k)}$ and $\olsi{\boldsymbol{F}}^{(k)}$ by inverting Eq.~\eqref{eq:locchark}.
	\item Treat the $k=0$ case specially as done in the example surrounding Eq.~\eqref{eq:surf0}, thereby obtaining $\olsi{\boldsymbol{q}}^{(0)}$ and by extension $\olsi{\boldsymbol{F}}^{(0)}$.\footnote{One could obviously use the safe variables and their derivatives, $\olsi{\boldsymbol{q}}^{(k)}$, to obtain the fluxes and their derivatives, $\olsi{\boldsymbol{F}}^{(k)}$, using analytic derivative formulae for $\boldsymbol{F}$. But those formulae are very cumbersome for high-order derivatives, so we find it easier to extrapolate the fluxes separately.}
	\item Solve for $\olsi{\boldsymbol{q}}^{(k)}$ and $\olsi{\boldsymbol{F}}^{(k)}$ for $k=1,2,3,4$ from $^{\boldsymbol{q}}\olsi{\boldsymbol{V}}^{*(k)}$ and $^{\boldsymbol{F}}\olsi{\boldsymbol{V}}^{*(k)}$ by inverting Eq.~\eqref{eq:locchark}.
	\item Populate split fluxes $\boldsymbol{f}_{\pm}$ wherever they are needed, for example $X_{\mathrm{ghost}}$.
	\item If needed, populate $\boldsymbol{q}_i$ at any interior points where values do not exist or must be replaced (see the discussion at the beginning of this section regarding the time update procedure, and see Sec.~\ref{app:fail} for policies to deal with failure modes).
\end{enumerate}

The surface is evolved by solving the equation $\partial_t x_v = \olsi{\alpha} \olsi{v}$ with a time-stepping algorithm (e.g.~Runge-Kutta), using the values of $\olsi{v}$ obtained above at each time sub-step. Note the right-hand side is the advective velocity of the surface $(u^x/u^t)|_{x_v}$, which is the actual velocity one sees the surface move with on a fixed grid. In this work, the lapse function is known analytically, but in fully general relativistic simulations one would have to interpolate it to the surface location with an order of accuracy that is consistent with the rest of the methods.

%
%%
%%%
%%%%
%%%
%%
%
\section{Implementation} \label{sec:imp}
In this section we present our numerical discretization. In Appendix~\ref{app:fail} we describe various failure modes that we have encountered, together with policies for dealing with them.

We use a uniform grid over a domain of length $L$ with grid spacing $\Delta x$. The system of equations of motion in semi-discrete form appear schematically in a flux-conservative form as
\begin{eqnarray}
\partial_t \boldsymbol{U}_i = - \frac{1}{\Delta x} \left( \boldsymbol{F}_{i+1/2} - \boldsymbol{F}_{i-1/2} \right) + \boldsymbol{S}_i .\label{eq:sys1}
\end{eqnarray}
The evolved variables are represented by $\boldsymbol{U}$, the fluxes are represented by $\boldsymbol{F}$, and any source terms that may be present are represented by $\boldsymbol{S}$. We discretize spatial derivatives using 5th- or 7th-order upwind-biased finite differences, together with a global Lax Friedrichs flux split (see e.g.~\cite{shu1998essentially}). The flux split is $\boldsymbol{F} = \boldsymbol{f}^{+} + \boldsymbol{f}^{-}$ where
\begin{eqnarray}
\boldsymbol{f}^{\pm} = \frac{1}{2} \left( \boldsymbol{F} \pm \mathrm{max}(|\lambda |) \boldsymbol{U}\right).
\end{eqnarray}
The eigenvalues of the flux Jacobian are represented by $\lambda$, and the maximum is taken over the entire grid. By adding such terms $\pm \mathrm{max}(|\lambda |) \boldsymbol{U}$, the eigenvalues of the Jacobians of $\boldsymbol{f}^{\pm}$ are respectively semi-positive and semi-negative, allowing us to separately apply the appropriate upwind finite difference operator  to each split flux.
Given values of the flux $\boldsymbol{f}^{\pm}$ at grid points (integer index $i$), 5th- and 7th-order left-biased approximations for $f^{+}_{i+1/2}$ and right-biased approximations for $f^{-}_{i-1/2}$~\cite{guercilena2017entropy} are given by
\begin{eqnarray}
^{5}\boldsymbol{f}^{\pm}_{i\pm 1/2} &=& \frac{1}{60}( 2\, \boldsymbol{f}^{\pm}_{i\mp 2} - 13\,\boldsymbol{f}^{\pm}_{i\mp 1} + 47\, \boldsymbol{f}^{\pm}_i + 27\,\boldsymbol{f}^{\pm}_{i\pm 1} - 3\, \boldsymbol{f}^{\pm}_{i\pm 2} ) \label{eq:FD5pm}\\
^{7}\boldsymbol{f}^{\pm}_{i\pm 1/2} &=& -\frac{1}{140}\boldsymbol{f}^{\pm}_{i\mp 3} + \frac{5}{84}\,\boldsymbol{f}^{\pm}_{i\mp 2} - \frac{101}{420}\boldsymbol{f}^{\pm}_{i\mp 1}+ \frac{319}{420}\boldsymbol{f}^{\pm}_i + \frac{107}{210}\boldsymbol{f}^{\pm}_{i\pm 1} - \frac{19}{210}\boldsymbol{f}^{\pm}_{i\pm 2} + \frac{1}{105}\boldsymbol{f}^{\pm}_{i\pm 3}. \label{eq:FD7pm} 
\end{eqnarray}
The discretization~\eqref{eq:sys1} thus reads
\begin{eqnarray}
\partial_t \boldsymbol{U}_i = - \frac{1}{\Delta x} \left( ^{s}\!\boldsymbol{f}^{+}_{i+1/2} +\, ^{s}\!\boldsymbol{f}^{-}_{i+1/2} -\, ^{s}\! \boldsymbol{f}^{+}_{i-1/2} -\, ^{s}\! \boldsymbol{f}^{-}_{i-1/2} \right) + \boldsymbol{S}_i, \label{eq:sys2}
\end{eqnarray}
where $s$ is the order of accuracy of the flux.

For time stepping we use a total variation diminishing (TVD) Runge-Kutta scheme of 3rd-order accuracy. For a system written schematically as $\partial_t U = \mathcal{L}(U)$, the time stepping reads
\begin{eqnarray}
U^{\{1\}} &=& U^n + \Delta t \mathcal{L}(U^n) \nonumber\\
U^{\{2\}} &=& \frac{3}{4} U^n + \frac{1}{4}\left( U^{\{1\}} + \Delta t \mathcal{L}(U^{\{1\}}) \right) \\
U^{n+1}   &=& \frac{1}{3}U^n + \frac{2}{3}\left( U^{\{2\}} + \Delta t \mathcal{L}(U^{\{2\}})\right),
\end{eqnarray}
where $(U^n,U^{n+1})$ are respectively the values of $U$ at time levels $(n,n+1)$, and $(U^{\{1\}},U^{\{2\}})$ are values at intermediate steps. In order to obtain $s$th-order accuracy in time, we use a time step $\Delta t \propto \Delta x^{s/3}$~\cite{tan2010inverse}. The surface position is evolved according to $\partial_t x_v = (\alpha v)|_{x_v}$ using the same Runge-Kutta time stepping.

We compare our high-order methods with a standard finite volume scheme using the Harten-Lax-van Leer (HLL) flux formula~\cite{harten1983upstream} and a standard artificial atmosphere with constant density $10^{-13}$ and a standard polytropic equation of state $P=100 \rho^2$ everywhere. We use the barotropic Valencia scheme, which is described in~\cite{westernacher2019hamilton}.
%
%%
%%%
%%%%
%%%
%%
%
\section{Numerical Results}\label{sec:results}

In what follows, we use the Valencia formulation given by Eqs.~\eqref{eq:toyvalboth}. We comment on using the Hamiltonian formulation in Sec.~\ref{sec:resultsham}.

\begin{figure}
\centering
\includegraphics[width=0.9\textwidth]{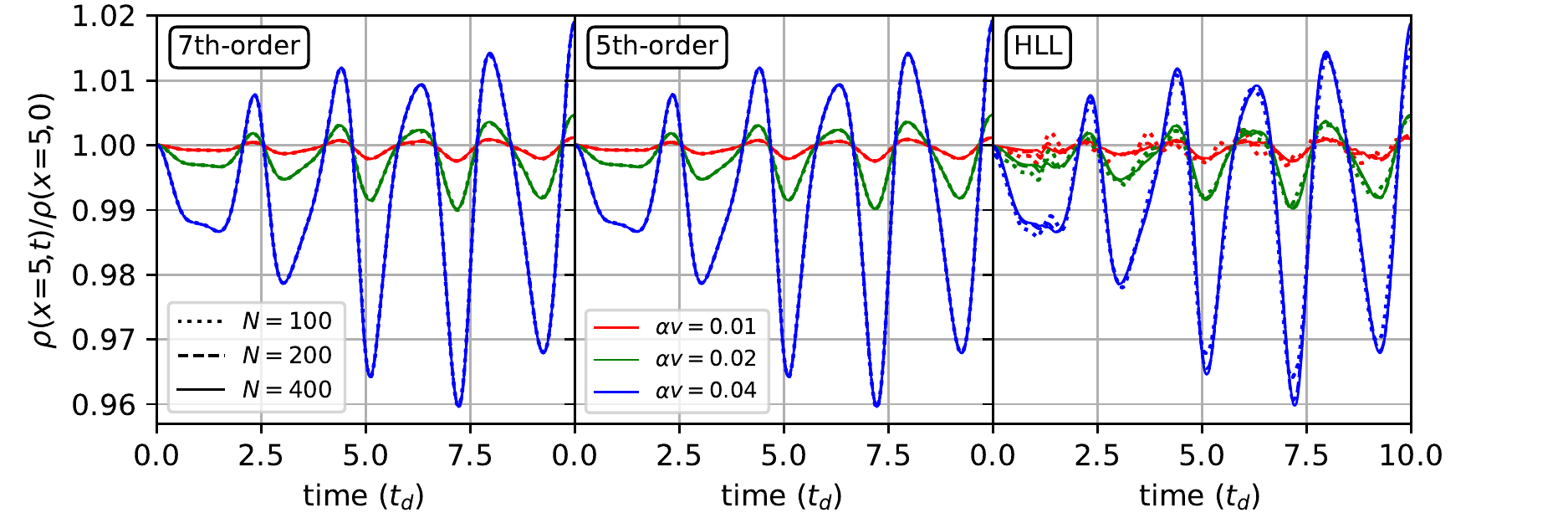}

\caption{Density evolution at the center of the grid $x=5$. Left, center, and right respectively show evolutions using the 7th-order, 5th-order, and HLL methods, all with CFL factor $1.0$. Three resolutions $N=\lbrace 101,201,401\rbrace$ are displayed using different line styles, and three initial uniform advective velocities $\alpha v = \lbrace 0.01, 0.02, 0.04 \rbrace$ are displayed using different colors. The dynamical time of the star is taken to be exactly $t_d=10$.} \label{fig:rhovst}
\end{figure}

\begin{figure}
\centering
\includegraphics[width=0.8\textwidth]{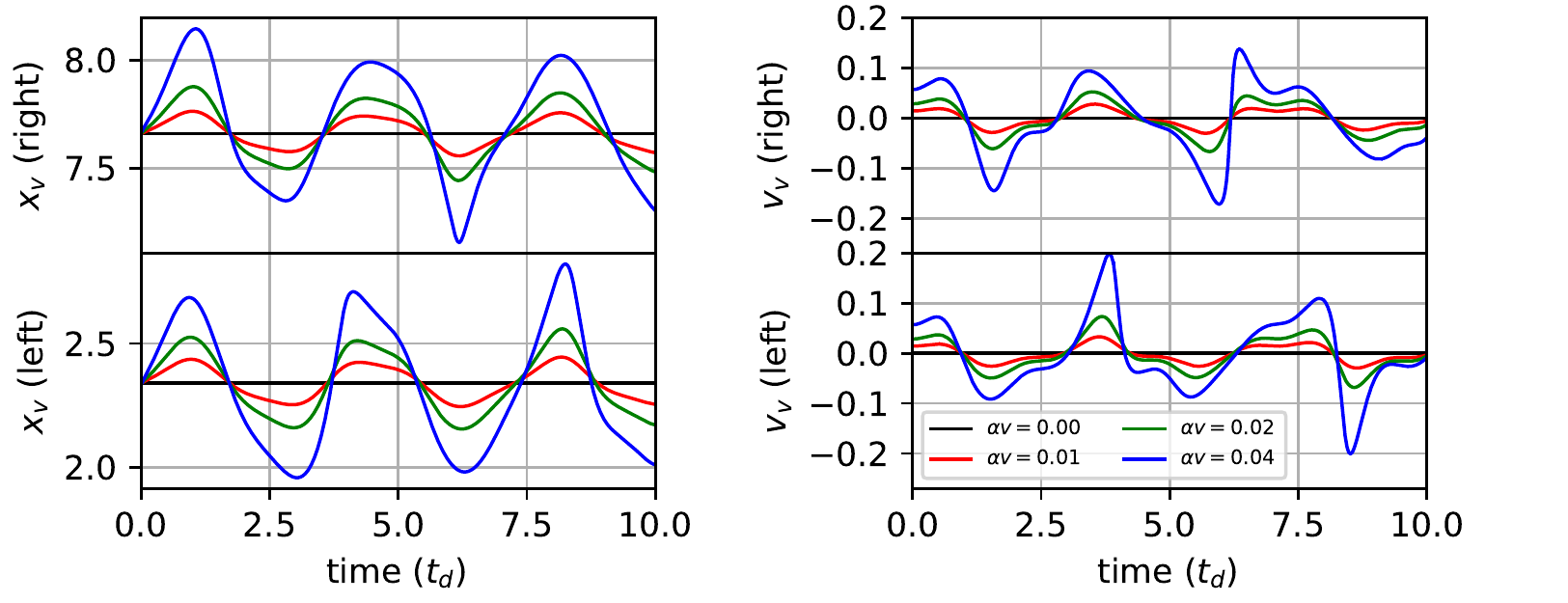}

\caption{Left and right surface positions (left column) and surface velocities (right column) over time, for resolution $N=400$ using the 7th-order method with Courant factor $1.0$. Cases with four initial uniform advective velocities $\alpha v = \lbrace 0.00, 0.01, 0.02, 0.04 \rbrace$ are displayed using different colors. The dynamical time of the star is taken to be exactly $t_d=10$.} \label{fig:surfvst}
\end{figure}

Some properties of our simulations are displayed in Figs.~\ref{fig:rhovst} \&~\ref{fig:surfvst}, which evolve for a total duration of 10 dynamical times $t_d$ (where we take $t_d$ to be exactly 10, although a more precise value according to the central energy density is $t_d = 1/\sqrt{\epsilon_c} \approx 10.72$). Fig.~\ref{fig:rhovst} shows the evolution of the rest mass density at the center of the grid, $x=5$, on a fractional basis relative to its initial value. The HLL method with artificial atmosphere (right) is shown along with our finite difference 5th-order (middle) and 7th-order (left) methods with surface tracking. Cases with three initial advective velocities $\alpha v \in \lbrace 0.01,0.02,0.04\rbrace$ are displayed using different colors, while we omit the equilibrium case $\alpha v = 0$ because it appears uneventful in this plot and clutters the visual inspection. Three different resolutions corresponding to a number of grid points across the whole domain of $N\in \lbrace 100,200,400\rbrace$ are shown using different line styles. The star initially spans roughly 53\% of the domain, and its span fluctuates as it sloshes around in the fixed spacetime. As the initial velocity is increased, the density oscillations increase correspondingly. It is qualitatively evident that the HLL method with artificial atmosphere (right) is slower to converge, because the curves for different resolutions do not overlap well in comparison to the surface-tracking methods (left and middle). The disparity in convergence and absolute level of error are presented quantitatively in several figures described below.

For the surface-tracking method, the evolution history of the surface positions and velocities are available. These are shown in Fig.~\ref{fig:surfvst}, where surface positions $x_v$ are displayed in the left column and surface velocities $v_v$ are displayed in the right column.\footnote{Note that the actual surface motion would be given by the advective velocity, $\alpha|_{x_v} v_v$.} We only display the 7th-order method in this figure, but we display cases with all initial velocities $\alpha v \in \lbrace 0.00, 0.01, 0.02, 0.04 \rbrace$. In Fig.~\ref{fig:timeresvst} below, which shows numerical errors over time, spikes in the numerical error correspond to maximal compression of the left and right sides of the star (since this yields maximal gradients at the surfaces). For the right surface, maximal gradients occur when $x_v$ reaches locally minimal values, whereas for the left surface, maximal gradients occur when $x_v$ reaches locally maximal values (compare Fig.~\ref{fig:timeresvst} to the left column of Fig.~\ref{fig:surfvst}).

\begin{figure}
\centering
\includegraphics[width=0.95\textwidth]{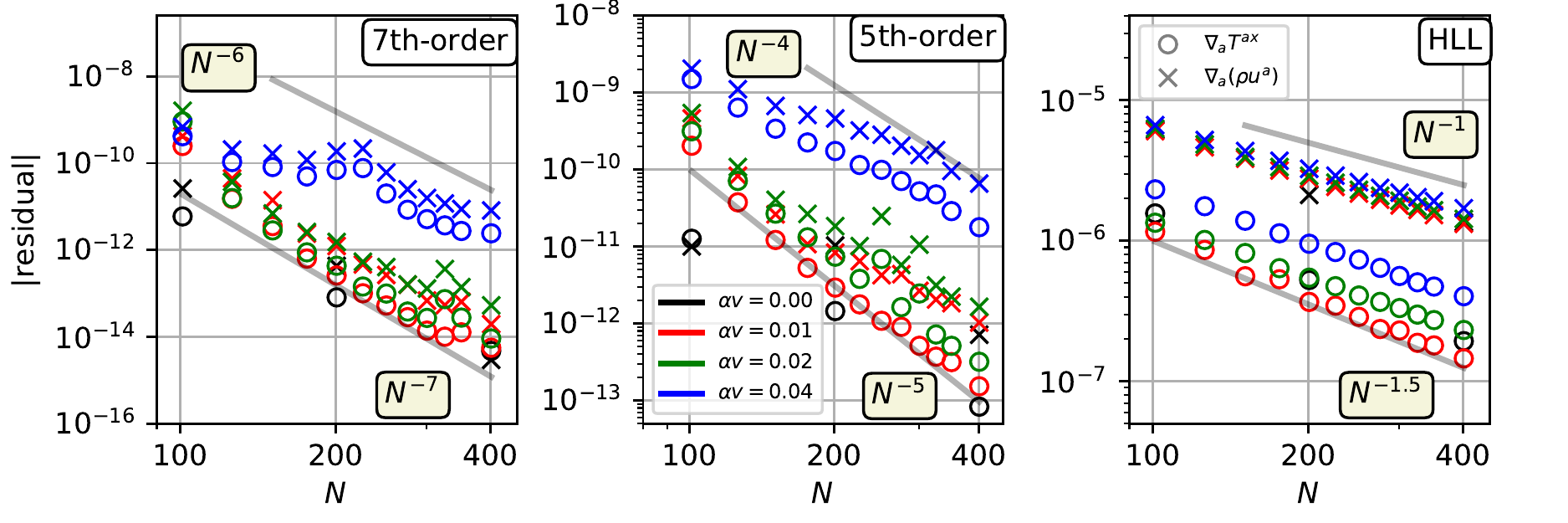}

\caption{Global convergence rates for the 7th-order scheme (left), 5th-order scheme (middle), and standard finite volume scheme using the HLL flux and an artificial atmosphere (right). Cases with four initial uniform advective velocities $\alpha v = \lbrace 0.00, 0.01, 0.02, 0.04 \rbrace$ are displayed using different colors. The momentum and mass residuals are displayed with circle and cross tick marks. The residuals are averaged over $\sim 10$ dynamical times and over the interior of the star. Power-law trends are indicated with grey lines. Comparing the three methods with each other is not meaningful in this figure, except for the convergence rates, because different time steps are used and the residuals are evaluated with different order finite difference operators (8th, 6th, and 4th, respectively for left, middle, and right). For the equilibrium star $\alpha v = 0$, we only display $N\in \lbrace 100,200,400 \rbrace$ for reasons explained in the text.} \label{fig:fullavgvsN}
\end{figure}

Next we present the convergence properties of the different methods on a quantitative basis. We quantify numerical errors by evaluating the residuals of the equations of motion, namely the local conservation of rest mass $|\nabla_a (\rho u^a)|$ and momentum $|\nabla_a T^{ai}|$ (both of which would be zero given analytic solutions). The numerically-obtained fluid variables are plugged into these residuals, and derivatives are computed using centered finite difference operators with an order of accuracy greater than the expected order of convergence (so as not to contaminate the measurement of convergence with the errors associated with such derivative evaluations). At the edges of the star, the residuals are not evaluated because the spatial derivative stencils extend into the vacuum regions. Similarly for the time derivatives at the temporal beginning and end of the simulations. We present the residuals on a space-and-time integrated basis versus resolution in Fig.~\ref{fig:fullavgvsN}, as well as on a space integrated basis versus time in Fig.~\ref{fig:timeresvst}. Generally speaking, we present a resolution range corresponding to $N\in [100,400]$. $N=100$ is slightly outside of the convergent regime for some initial velocities, whereas machine precision limits our ability to measure convergence for the surface-tracking methods when $N \gtrsim 400$.

In Figs.~\ref{fig:fullavgvsN} \&~\ref{fig:timeresvst}, the finite difference orders used to evaluate the residuals for the 7th-order, 5th-order, and HLL methods are respectively 8th-order, 6th-order, and 4th-order. The time steps used are also not the same. In all methods, the Courant-Friedrichs-Lewy factor is $1$ for the lowest resolution corresponding to $N=100$. But for the 7th- and 5th-order cases, the Courant factor is respectively $(\Delta x/0.1)^{7/3-1}$ and $(\Delta x/0.1)^{5/3-1}$, where $\Delta x$ is the spatial resolution $\Delta x = L/N$ (these choices maintain 7th- and 5th-order accuracy in time, even though we use a 3rd-order accurate Runge-Kutta time-stepping method). For these reasons, it is not fair to compare the absolute levels of error between the three methods in Figs.~\ref{fig:fullavgvsN} \&~\ref{fig:timeresvst}. However, it is fair to compare the convergence orders.

Note that it is more delicate to measure convergence in the case of an equilibrium star, $\alpha v = 0$, because as grid points are added to the domain, the distance between the outermost interior points and the respective surfaces does not decrease monotonically, which then reflects upon the residuals as large fluctuations versus $N$.\footnote{Although if one appropriately shifts the position of the star with respect to the grid as points are added, then the distance between the outermost interior points and the respective surfaces would decrease monotonically.} For this reason, we display resolutions which are only integer multiples of $N=100$, namely $N\in \lbrace 100,200,400\rbrace$, which yields meaningful measurements of convergence because grid point positions are never removed as $N$ doubles, only added. In non-equilibrium cases $\alpha v \neq 0$, this issue is less pronounced because the surface moves across grid points anyway; the numerical error associated with the initial position of the star on the grid is less dominant than the errors associated with the stellar motion. We believe this concern is a peculiarity of the 1-dimensional case because there are only 2 outermost interior points. In multiple dimensions, there will be many outermost interior points with a well-populated range of shortest distances to the surface, thereby contributing a diverse range of numerical errors to the total error budget. In aggregate, such errors would not fluctuate drastically as the position of the star shifts slightly with respect to increasingly larger grids.

In Fig.~\ref{fig:fullavgvsN}, the mass and momentum residuals are displayed using different tick marks, cases with different initial advective velocities $\alpha v \in \lbrace 0.00,0.01,0.02,0.04\rbrace$ are displayed using different colors, and grey solid lines are reference power-law trends. We observe 7th- and 5th-order convergence (left and middle) for low values of the initial advective velocity $\alpha v$. For the most extreme case with $\alpha v = 0.04$, the convergence order degrades by roughly 1, and it becomes apparent that the convergent regime is roughly $N\geq 200$. By contrast, the standard HLL method with artificial atmosphere yields the widely observed $1.5$ convergence order, with a noticeable degradation to $1$ for the mass residual. The degradation of the mass residual is also noticeable for the 5th-order method, but appears less pronounced for the 7th-order method. The explanation for this is unclear.

In Fig.~\ref{fig:timeresvst} we show the spatially integrated residuals as a function of time. The mass and momentum residuals are displayed using different line styles, and a selection of three resolutions corresponding to $N \in \lbrace 125, 200, 400\rbrace$ are displayed using different colors. Only the most extreme initial velocity is shown, $\alpha v = 0.04$, because this case most clearly illustrates the spiking of numerical errors coinciding with the local maxima of surface gradients (compare with Fig.~\ref{fig:surfvst}). In some instances, for example comparing $N \in \lbrace 125,200\rbrace$ for the 7th-order method (left), the residual does not settle down to previous levels following a spike (look at time $\sim 6.25\, t_d$). This behavior, as well as the spiking in residuals themselves, are what account for the degradation of global convergence order observed in Fig.~\ref{fig:fullavgvsN} for the most extreme case $\alpha v = 0.04$. The residuals in the HLL method with artificial atmosphere also exhibit spikes at these times, although they are somewhat less pronounced.

\begin{figure}
\centering
\includegraphics[width=0.95\textwidth]{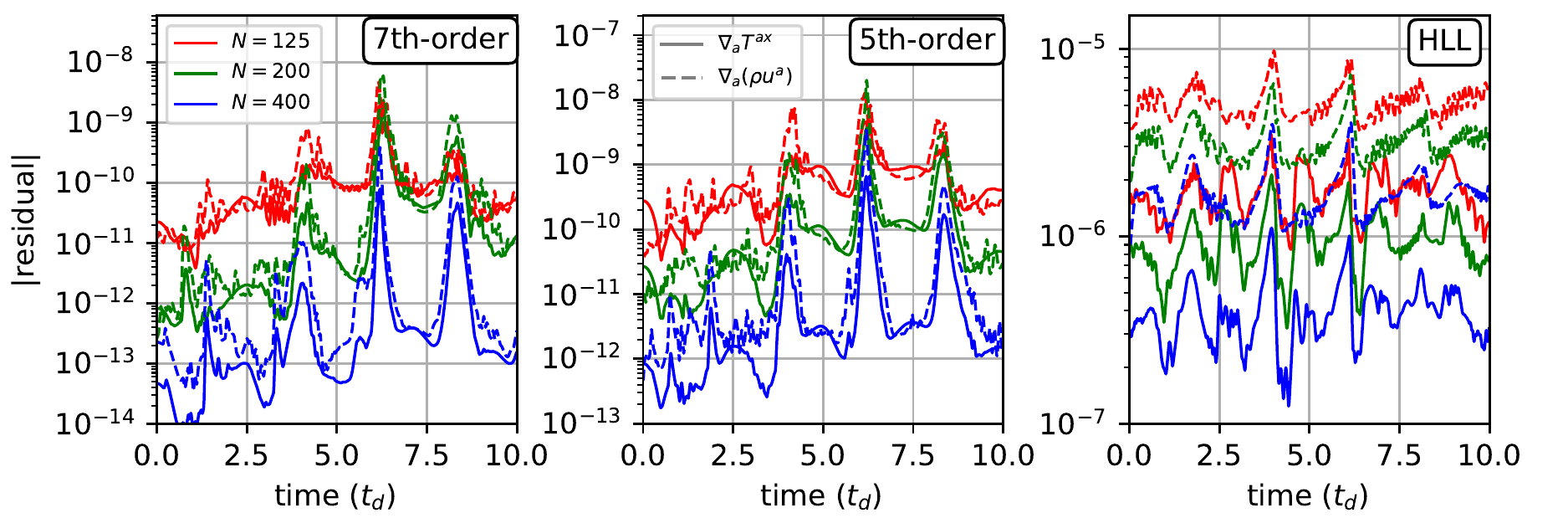}

\caption{Spatially-averaged residuals over time for the 7th-order scheme (left), 5th-order scheme (middle), and standard finite volume scheme using the HLL flux and an artificial atmosphere (right). Only the most extreme case with initial uniform advective velocity $\alpha v = 0.04$ is displayed, which exhibits the clearest spikes in residuals (and degradation of convergence order) when surface gradients and acceleration of the surface become large. A selection of resolutions is displayed, corresponding to a number of grid points $N=\lbrace 125,200,400\rbrace$. The momentum and mass residuals are displayed with solid and dashed lines. The residuals are averaged over the interior of the star, and the dynamical time is taken to be exactly $t_d = 10$. The curves have been gently smoothed with a Gaussian kernel of width $0.2\times t_d$, in order to improve the visual presentation. Caution should be used when comparing the three methods using this figure, because different time steps are used and the residuals are evaluated with different order finite difference operators (8th, 6th, and 4th, respectively for left, middle, and right).} \label{fig:timeresvst}
\end{figure}

\begin{figure}
\centering
\includegraphics[width=0.95\textwidth]{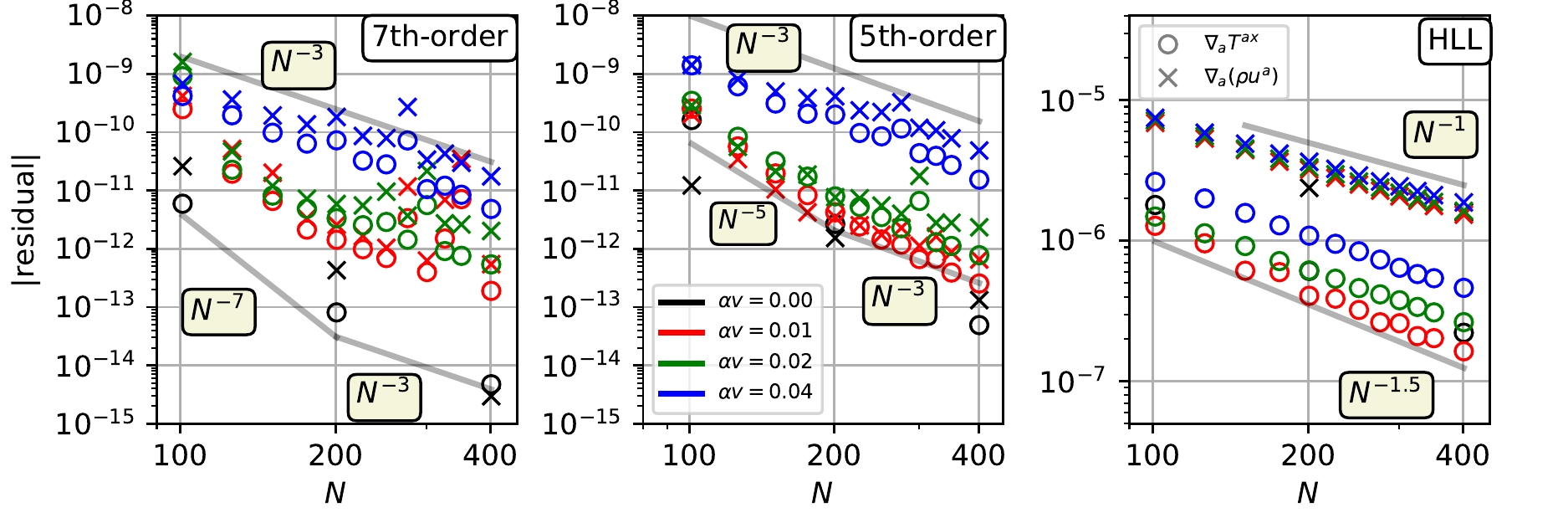}

\caption{Global convergence rates with a fixed Courant-Friedrichs-Levy factor of $1.0$ for the 7th-order scheme (left), 5th-order scheme (middle), and standard finite volume scheme using the HLL flux and an artificial atmosphere (right). Cases with four initial uniform advective velocities $\alpha v = \lbrace 0.00, 0.01, 0.02, 0.04 \rbrace$ are displayed using different colors. The momentum and mass residuals are displayed with circle and cross tick marks. The residuals are averaged over $\sim 10$ dynamical times and over the interior of the star. Power-law trends are indicated with grey lines. Comparing the three methods with each other is meaningful in this figure, because the same time stepping is used and the residuals are evaluated with the same (8th-order) finite difference operator. The convergence order for the 7th-order and 5th-order methods quickly drops to 3rd-order as time step errors become dominant, reflecting the 3rd-order Runge Kutta time-stepping method. For the equilibrium star $\alpha v = 0$, we only display $N\in \lbrace 100,200,400 \rbrace$ for reasons explained in the text.} \label{fig:apples}
\end{figure}

In order to make a fair comparison between the different methods on the basis of absolute level of residuals, in Fig.~\ref{fig:apples} we repeat the same style of plots as in Fig.~\ref{fig:fullavgvsN}, except we use the same Courant factor of $1$ across all cases and we use the same order finite difference operator (8th-order) when evaluating all residuals. The absolute level of error as measured by these residuals is roughly 5 to 6 times smaller for the surface-tracking methods than for the HLL method with artificial atmosphere. Furthermore, for the surface-tracking methods, we observe that the convergence order degrades to 3 as one moves either to larger $\alpha v$ or larger $N$. This reflects the dominance of the numerical error associated with time stepping, which with fixed Courant factor decreases as 3rd order for our Runge-Kutta method. This figure shows that one does not necessarily need a time-stepping method higher than 3rd order to achieve the benefits of low numerical error. However, higher order strong stability preserving time-stepping methods exist~\cite{gottlieb2009high} which could be utilized to achieve an overall convergence order of 5 or 7, for example.

\begin{figure}
\centering
\includegraphics[width=0.9\textwidth]{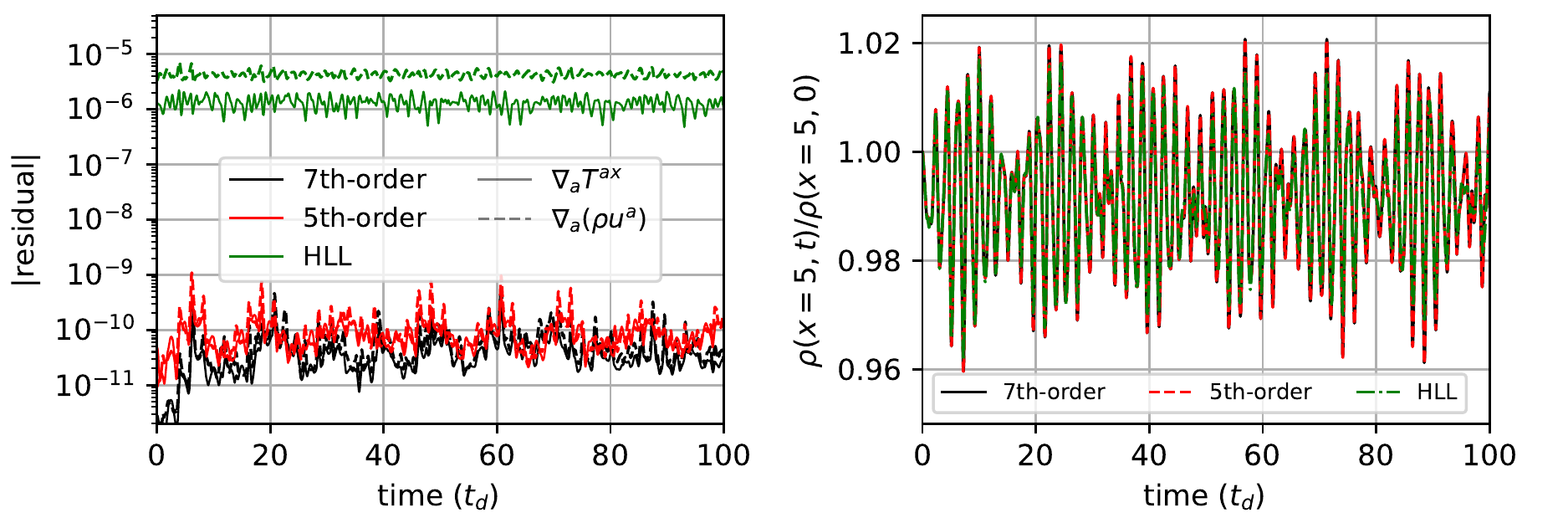}

\caption{This figure shows long-term stability over 100 dynamical times. Spatially-averaged residuals over time for the 7th-order, 5th-order, and standard finite volume scheme using the HLL flux and an artificial atmosphere (left). Only the most extreme case with initial uniform advective velocity $\alpha v = 0.04$ is displayed. One resolution is displayed ($N=150$), corresponding to roughly the lowest resolution for which long-term stability is achieved using all methods. The momentum and mass residuals are displayed with solid and dashed lines. The residuals are averaged over the interior of the star, and the dynamical time is taken to be exactly $t_d = 10$. The curves have been gently smoothed with a Gaussian kernel of width $t_d$, in order to improve the visual presentation. Different time steps are used and the residuals are evaluated with different order finite difference operators (8th, 6th, and 4th, respectively for 7th-order, 5th-order, and HLL methods). Central density evolutions from the same simulations are displayed on the right.} \label{fig:stab}
\end{figure}

Lastly, we demonstrate long-term stability of each method in Fig.~\ref{fig:stab}. We show only the most extreme case $\alpha v = 0.04$, which is the most demanding test for long-term stability. We use a single resolution corresponding to $N=150$, which is roughly the lowest number of grid points which yields stable evolution over 100 dynamical times. The left plot shows the mass and momentum residuals over time, and stability is evidenced by the absence of any apparent secular drift in residuals. The right plot shows the rest mass density evolution at the center of the grid $x=5$, on a fractional basis with respect to its initial value. The HLL method with artificial atmosphere has a noticeable secular decay in central density oscillations relative to the surface-tracking 7th- and 5th-order methods, meaning that the latter methods are less dissipative at the same grid resolution. See supplemental material at [URL will be inserted by publisher] for a video of the long-term evolution.

\subsection{Hamiltonian Formulation} \label{sec:resultsham}
In this subsection we briefly comment on the possibility of using the Hamiltonian formulation with the surface-tracking methods. It was found in~\cite{westernacher2019hamilton} that surface stability issues were a persistent problem when using the HLL method with an artificial atmosphere, and this problem was solved by using a hybrid scheme whereby the Valencia formulation is used at the last interior point of the star and beyond. Our tests in this work indicate that it is possible to use the pure Hamiltonian formulation, and the convergence properties are similar to the Valencia formulation. However, the stability is greatly diminished. Failure modes are encountered more often, requiring the invocation of error handling policies, whereas with the Valencia formulation we almost never encounter a failure mode (unless the initial velocity of the star is roughly $\alpha v = 0.6$ or greater, and even then one expects failure modes to be avoided with a shock-capturing scheme since large surface gradients would be more well-tolerated). We were not able to evolve any moving star cases stably for greater than about 30 dynamical times with the Hamiltonian formulation. The stability issues are again surface-related, which probably stem from the fact that the weak solutions for a fluid-vacuum interface in the Hamiltonian formulation support shockwaves~\cite{westernacher2020exploration} (whereas in the Valencia formulation only rarefaction waves are supported~\cite{munz1994tracking, toro2013riemann}). Thus, in future applications of the Hamiltonian formulation with our surface-tracking methods, we expect it to be advisable and perhaps necessary to once again hybridize the formulation with Valencia at the outermost interior points, as done in Ref.~\cite{westernacher2019hamilton}.
%
%%
%%%
%%%%
%%%
%%
%

\section{Summary} \label{sec:summary}
Future gravitational wave detections will have signal-to-noise ratios which challenge the accuracy of current state-of-the-art waveform models calibrated to numerical relativity simulations. In numerical relativity simulations of compact binaries involving at least one material body, the standard way that stellar surfaces are treated (using an artificial atmosphere) is a significant source of error and the primary obstacle preventing high-order convergence during the inspiral phase of the binary. We presented a new numerical method which tracks and evolves the stellar surface, and imposes free surface boundary conditions there with an arbitrarily high order of accuracy (in principle). Our surface-tracking method is independent of the scheme used in the interior of the star, which should allow wide adoption of our method. We demonstrated this new method in a 1+1-dimensional proof-of-principle numerical implementation, whereby a relativistic 1-dimensional star with a barotropic liquid equation of state undergoes violent sloshing motion inside of a fixed gravitational potential. The method performs very well in comparison to a standard finite volume scheme with an artificial atmosphere, both in convergence rate as well as a decrease in absolute numerical errors by 5-6 orders of magnitude at the same resolution. In the future, we anticipate that this method will be applied to simulate the inspiral phase of material compact binaries in full 3+1-dimensional numerical relativity (which is expected to be barotropic fluid flow to a high degree of accuracy).

A challenge for multiple dimensions, which we have not addressed in this work, is dealing not just with a moving boundary (the surface), but with a deformable one. We anticipate that non-oscillatory interpolations of the surface position will be necessary to prevent excessive unphysical wrinkling of the stellar surface in multiple dimensions. However, this is not merely a technical challenge; it is also an opportunity to input more precise surface tension physics into numerical simulations, which may be accessible observationally with future gravitational wave detectors.

We do not believe that extending our surface-tracking method to baroclinic liquids would be a significant challenge, and there are no obvious impediments to using a realistic, tabulated equation of state. However, extending our method to the merger phase of compact binaries, whereby matter is being stripped and ejected, would be very challenging. From a geometrical standpoint, the number of distinct fluid-vacuum interfaces during the merger phase would undergo a rapid bifurcation cascade, with each surface undergoing potentially very large shape variations, which would require an exceptionally robust surface-tracking algorithm.

%
%%
%%%
%%%%
%%%
%%
%
\acknowledgements
We thank Charalampos Markakis, Vasileios Paschalidis, Frans Pretorius, Sirui Tan, and Chi-Wang Shu for discussions, as well as Nikolaos Tsakiris and Nick Stergioulas for inspiration~\cite{tsakiris2007vactrack} on using a surface-tracking algorithm. This research is supported by National Science Foundation Grant No. PHY-1912619 at the University of Arizona.

\bibliography{fluidbib}

%
%%
%%%
%%%%
%%%
%%
%
\appendix

\section{Surface Derivative Constraint in the Toy Model} \label{app:surfderconstrainttoy}
Since the fluid flow in the vicinity of the surface matches the surface motion, any quantity that takes on a persistent value at the stellar surface will have a vanishing Lagrangian time derivative there $D/Dt = \partial_t + (u^x/u^t)\partial_x$~\cite{baumgarte2010numerical}, where in our toy model $u^x/u^t=\alpha v$. For example, the pressure is persistently zero at the surface, and therefore satisfies
\begin{eqnarray}
\olsi{\frac{D P}{Dt}} = \olsi{\partial_t P} + \olsi{\alpha} \olsi{v} \olsi{\partial_x P} = 0
\end{eqnarray}
at the surface. The density is persistently equal to $\bar{\rho}$, defined by $P(\bar{\rho})=0$, thus we also have $\olsi{D\rho/Dt}=0$. Similarly, we have $\olsi{D(\rho h)/Dt}=0$ at the surface.

We can derive a condition on the spatial derivatives of the fluid variables at the surface using these vanishing Lagrangian time derivatives and the fluid equations. Write the rest mass conservation Eq.~\eqref{eq:toyrest} in terms of the Lagrangian time derivative,
\begin{eqnarray}
0 &=& \partial_t(\rho W) + \partial_x \left( \alpha \rho W v \right) \nonumber\\
&=& \frac{D}{Dt}(\rho W) + \rho W \partial_x \left( \alpha v \right) \nonumber\\
&=& \rho \frac{DW}{Dt} + W \frac{D\rho}{Dt} + \rho W \partial_x (\alpha v).
\end{eqnarray}
Evaluation at the surface yields
\begin{eqnarray}
0 &=& \olsi{\rho} \olsi{\frac{DW}{Dt}} + \olsi{\rho}\olsi{W} \olsi{\partial_x(\alpha v)}. \label{eq:surfden}
\end{eqnarray}
Note that $DW/Dt = W^3 v Dv/Dt$. The energy equation $\nabla_a T^{at}$ for our toy model expressed in a 1+1 decomposition of spacetime is~\cite{Alcubierre:2008}
\begin{eqnarray}
0 &=& \partial_t\left(\rho h W^2 - P \right) + \partial_x \left( \alpha \rho h W^2 v\right) + \rho h W^2 v \partial_x \alpha \nonumber\\
&=& \partial_t\left(\rho h W^2\right) -\frac{DP}{Dt} + \alpha v \partial_x P + \partial_x \left( \alpha \rho h W^2 v\right) + \rho h W^2 v \partial_x \alpha,
\end{eqnarray}
where in the second line we made the replacement $-\partial_t P = -DP/Dt +\alpha v \partial_x P$. Writing the remaining Eulerian time derivatives in terms of Lagrangian ones, we obtain
\begin{eqnarray}
0 &=& W^2\frac{D\left(\rho h\right)}{Dt} + \rho h 2 W \frac{DW}{Dt} -\frac{DP}{Dt} + \alpha v \partial_x P + \rho h W^2 \partial_x \left( \alpha v\right) + \rho h W^2 v \partial_x \alpha.
\end{eqnarray}
Evaluation at the surface causes $D(\rho h)/Dt$ and $DP/Dt$ to vanish and $\olsi{h}=1$:
\begin{eqnarray}
0 &=& \olsi{\rho}2\olsi{W} \olsi{\frac{DW}{Dt}} + \olsi{\alpha} \olsi{v} \olsi{\partial_x P} + \olsi{\rho}\olsi{W}^2\olsi{\partial_x\left( \alpha v \right)} + \olsi{\rho} \olsi{W}^2 \olsi{v} \olsi{\partial_x \alpha }. \label{eq:surfen}
\end{eqnarray}
Eliminating $\olsi{DW/Dt}$ using Eqs.~\eqref{eq:surfen} \&~\eqref{eq:surfden} yields the constraint on spatial derivatives of the fluid variables:
\begin{eqnarray}
\olsi{v} \olsi{\partial_x P} = \olsi{\rho} \olsi{W}^2 \olsi{\partial_x v}. \label{eq:surfderconstraint}
\end{eqnarray}

This constraint Eq.~\eqref{eq:surfderconstraint} must be satisfied at a stellar surface satisfying the boundary conditions~\eqref{eq:surfbcs}. Note we can rewrite $\partial_x v$ in terms of the primitive-like variable $\partial_x (Wv)$ via $\partial_x (Wv)=W^3 \partial_x v $. To gain an intuition for what the constraint Eq~\eqref{eq:surfderconstraint} means, note first that for an equilibrium star ($\partial_x (\alpha v)=0$) in the rest frame $v=0$, $\olsi{\partial_x P}$ is left unconstrained (as one would expect, since there are many possible equilibrium stars with varying $\olsi{\partial_x P}$). For an equilibrium star ($\partial_x (\alpha v)=0$) in a boosted frame $v\neq 0$, Eq.~\eqref{eq:surfderconstraint} expresses the condition of hydrostatic equilibrium at the surface in that frame. 

An interesting implication of Eq.~\eqref{eq:surfderconstraint} is that one cannot initialize a star in an arbitrary but seemingly reasonable manner and expect it to have a free surface. For example, initializing an equilibrium density profile with $\olsi{\partial_x P}\neq 0$ but setting $v=\mathrm{constant}>0$ results in a violation of Eq.~\eqref{eq:surfderconstraint}. In order for the initial data to satisfy the free surface vacuum boundary conditions~\eqref{eq:surfbcs}, care must be taken to enforce Eq.~\eqref{eq:surfderconstraint} as well. Combining instead rest mass conservation $\nabla_a (\rho u^a)=0$ with the Euler equation $\nabla_a T^a_i=0$ yields the same constraint as Eq.~\eqref{eq:surfderconstraint}.
%
%%
%%%
%%
%

\section{Inverse-Lax Wendroff Procedure for First Derivatives} \label{app:ILW}

Denote $k$th-order spatial derivatives of quantities with a superscript, e.g.~$\boldsymbol{q}^{(k)}$. Note that the matrix of left eigenvectors in Eq.~\eqref{eq:locchar0} is evaluated at the point $x_R$, whereas $\boldsymbol{V}$ and $\boldsymbol{q}$ are evaluated at arbitrary points $x_i$. Thus, derivatives of the fluid variables also obey this relation because the derivative operator treats the matrix of left eigenvalues at $x_R$ as constant. At the surface, using the safe variables from Sec.~\ref{sec:toyA}, we obtain
\begin{eqnarray}
\olsi{\boldsymbol{V}}^{(k)} =
	\begin{bmatrix}
	\phantom{+}W/(h c_s) & 1 \\
	-W/(h c_s) & 1 
%%%% for (\rho,Wv) formulation
%	\phantom{+}Wc_s/\rho & 1 \\
%	-Wc_s/\rho & 1 
\end{bmatrix}_R
\olsi{\boldsymbol{q}}^{(k)}. \label{eq:locchar}
\end{eqnarray}
With extrapolation of the outgoing first-order derivative, we have
\begin{eqnarray}
\olsi{V}^{*(1)}_{+} = \left(\frac{W}{h c_s}\right)_{\!\! R} \olsi{\partial_x h} + \olsi{\partial_x(Wv)}, \label{eq:surf1temp}
\end{eqnarray}
where there are two unknowns $\olsi{\partial_x h}$ and $\olsi{\partial_x (Wv)}$. We can eliminate one unknown using the constraint Eq.~\eqref{eq:surfderconstraint}, which we write in terms of $\olsi{\partial_x \rho} = \olsi{\partial_h \rho} \olsi{\partial_x h} = (\olsi{\rho}/\olsi{c}_s^2)\olsi{\partial_x h}$ and $\olsi{\partial_x (Wv)}$ as
\begin{eqnarray}
\olsi{W}\olsi{v} \olsi{\partial_x h} = \olsi{\partial_x Wv}. \label{eq:surfder2}
\end{eqnarray}
Plugging into Eq.~\eqref{eq:surf1temp} gives
\begin{eqnarray}
\olsi{V}^{*(1)}_{+} = \left[\left(\frac{W}{h c_s}\right)_{\!\! R} + \olsi{W}\olsi{v} \right] \olsi{\partial_x h}, \label{eq:surf1}
\end{eqnarray}
which we can use to solve for $\olsi{\partial_x h}$ and then recover $\olsi{\partial_x (Wv)}$ using Eq.~\eqref{eq:surfder2}. This expression is not invertible for $\olsi{\partial_x h}$ when $\left(\frac{W}{h c_s}\right)_{\!\! R} + \olsi{W}\olsi{v}=0$. In this unlikely scenario, one can instead extrapolate both characteristic variables to obtain $V^{* (1)}_{\pm}$, and then invert Eq.~\eqref{eq:locchar} to obtain $\boldsymbol{q}^{*}$.

This would constitute the inverse Lax-Wendroff procedure for the first derivatives, but we find it to be unnecessary. For derivatives higher than first order, in principle one can derive constraints on higher-order spatial derivatives of the fluid variables following a similar strategy as we did for Eq.~\eqref{eq:surfderconstraint}. For example, one can use the fact that $\olsi{D^k P/Dt^k}=\olsi{D^k \rho/Dt^k}=\olsi{D^k \rho h/Dt^k}=0$. But this is numerically unnecessary, and with a dynamical spacetime would actually involve time derivatives of the metric, leading to very cumbersome constraints on the fluid derivatives at the surface.

\section{Failure Modes and Error-Handling Policies} \label{app:fail}
There are failure modes we have encountered during the development of our surface-tracking method. They all involve the density becoming negative at interior stellar points close to the surface.

In the event that the density at the last interior point becomes negative or zero, it is repopulated using an interpolation of the same order as the rest of the methods (e.g. 5th- or 7th-order). If the interpolated replacement also has a negative density, then the surface position is manually receded to some fraction $A$ of a grid separation $\Delta x$ away from the last interior positive density point. This failure mode might occur, for example, if the surface position is a tiny distance away from the last interior point, such that the density at the last interior point is very small.

If the density at multiple consecutive last interior points (e.g.~the last two points) becomes negative or zero, then the surface position is reset to some fraction of a grid separation $A \Delta x$ away from the last positive density point, and the points at which the density became negative are therefore deemed to be in the vacuum region. One could alternatively attempt to repopulate the points with physical values, but we interpret this failure mode as an indication that the surface evolution has become too inconsistent with the fluid.

Note that these failure modes almost never occur for the Valencia formulation. For the Hamiltonian formulation, these failure modes occur often, thus such evolutions are unsupportable without the error handling policies. When using the Valencia formulation, the failure modes above usually indicate an implementation bug. Generally speaking, the latter failure mode (negative density at multiple consecutive points) is more severe.

\end{document}